\documentclass{JHEP3} 

\usepackage[dvipdfm]{graphicx}
\usepackage{bm}
\usepackage{amsmath,amssymb}
\usepackage{multicol,bbm}
\usepackage{subfigure}
\newcommand\fverb{\setbox\fverbbox=\hbox\bgroup\verb}
\newcommand\fverbdo{\egroup\medskip\noindent%
            \fbox{\unhbox\fverbbox}\ }
\newcommand\fverbit{\egroup\item[\fbox{\unhbox\fverbbox}]}
\newbox\fverbbox
\newcommand{\be}{\begin{equation}}
\newcommand{\ee}{\end{equation}}
\newcommand{\bea}{\begin{eqnarray}}
\newcommand{\eea}{\end{eqnarray}}
\newcommand{\ba}{\begin{array}}
\newcommand{\ea}{\end{array}}

\newcommand{\nn}{\nonumber}

\title{Holographic phase transition of QCD with back-reaction of flavors }

\author{Youngman Kim
        \\
Asia Pacific Center
for Theoretical Physics and Department of Physics, Pohang University
of Science and Technology, Pohang, Gyeongbuk 790-784, Korea
\\  E-mail: \email{ykim@apctp.org}}
\author{ Tatsuhiro Misumi
\\
Yukawa Institute for Theoretical Physics, Kyoto University, Kyoto 606-8502, Japan
\\ E-mail: \email{misumi@yukawa.kyoto-u.ac.jp}}
\author{Ik Jae Shin
\\
Asia Pacific Center
for Theoretical Physics, Pohang, Gyeongbuk 790-784, Korea
\\ E-mail: \email{geniean@apctp.org}}
\received{\today}       

\abstract{
We revisit confinement/deconfinement transition in holographic QCD to consider the back-reaction
of a bulk scalar field.
The bulk scalar field is dual to a quark bi-linear operator $\bar qq$, and it
encodes explicit and spontaneous chiral symmetry breaking of QCD.
To perform the Hawking-Page transition analysis with the back-reaction,
we first obtain a deformed AdS black hole solution due to a finite quark mass.
Through the Hawking-Page analysis with the back-reacted geometry,
we study the flavor number $N_f$ and finite quark mass dependence of the critical temperature
of the QCD confinement/deconfinement transition. In addition, we confirm that unphysical QCD phase haunting around in a previous study disappears with the back-reaction.
We discuss our results in the light of lattice QCD observations.
}


\begin{document}

\section{Introduction}
Understanding the QCD phase diagram is one of key issues in modern theoretical physics;~see \cite{QCDphaseDreviews} for some recent reviews.
Studying every nook and corner of the phase diagram requires nonperturbative approaches to QCD.
AdS/QCD or holographic QCD based on the AdS/CFT~\cite{adscft} might be one of the promising tools for
strongly interacting regime of QCD.
Surely in the present form of holographic QCD, we are not to say  much about the QCD phase diagram.
For instance the nature of the QCD phase transition is intimately involved in the number of quark flavors and the quark mass:
it can be first order, second order, and crossover depending on them,
while in holographic QCD the transition is the first order due to the large $N_c$ nature of the Hawking-Page transition \cite{Witten}.
With this caveat, we take one more step to improve our understanding on the QCD phase diagram in holographic QCD.

The Hawking-Page transition in a sliced AdS was discussed in~\cite{Herzog}, and effects of matter, quarks or mesons, were worked out in~\cite{Kim:2007em} with no back-reaction.
In \cite{CO} $\alpha^\prime$ corrections are studied.
In \cite{Kim:2007em} it is observed that the effects of matter without back-reaction lead to an unphysical phase transition at low temperature,
which suggests that we need to take into account the back-reaction of the matter field to discuss the phase transition more precisely.
It is reasonable from the viewpoint of  the large $N_{c}$ correction since the contribution from the back-reaction and that from the matter part of the action
both belong to the same order of $1/N_{c}$ corrections.

In this paper, we study the confinement/deconfinement transition in holographic QCD considering the back-reaction of a bulk scalar field that is dual to a quark bi-linear operator $\bar qq$.
To perform the Hawking-Page transition analysis with the back-reaction,
we first obtain a deformed AdS black hole solution due to a finite quark mass.
Through the Hawking-Page analysis with the back-reacted geometry,
we calculate the flavor number $N_f$ and finite quark mass dependence of the critical temperature
of the confinement/deconfinement transition of QCD.
We also confirm that unphysical QCD phase observed in \cite{Kim:2007em} disappears with the back-reaction.

\section{Back-reacted AdS solutions}
In this section we consider AdS solutions deformed by the bulk scalar field.
The back-reacted thermal AdS (tAdS) solution in the hard wall model~\cite{EKSS, PR} has obtained in \cite{Shock}, while
a back-reacted AdS black hole (AdSBH) solution due to the bulk scalar field has not been discussed.
Thus we need to obtain the deformed AdSBH solution.
We begin with the following $5$-dimensional action,
\begin{equation}
S \,=\, \int d^{5}x \sqrt{g}~[-{1\over{2\kappa^{2}}}R\,+\,{\rm Tr}(\partial_{M} X\partial^{M}X\,-\,V(X))],
\label{S0}
\end{equation}
where $R$ is the $5$-dimensional Ricci scalar, $X$ is a scalar field and $\kappa^{2}=8\pi G_{5}$ and $1/G_{5}=32N^{2}_{c}/\pi L^{3}$ ($M=0,1,2,3,4$).
Here we assume that the scalar solutions are given by $X=X_{0}(z)=({\bf 1}_{N_{f}\times N_{f}}/2)(m_{q}z+\sigma z^{3})$ for tAdS \cite{Shock} and $X=X_{0}(z)=({\bf 1}_{N_{f}\times N_{f}}/2)m_{q}z$ for AdSBH where $N_{f}$ is the number of quark flavors.
In the hard wall model $m_{q}$ and $\sigma$ stand for the quark mass and the chiral condensate, up to a constant, respectively.
We take the AdS radius $L$ as one, $L=1$.
After taking a trace ${\rm Tr}$ the action is rewritten as follows
\begin{equation}
S \,=\, \int d^{5}x \sqrt{g}~[-{1\over{2\kappa^{2}}}R\,+\,{N_{f}\over{4}}\partial_{M} \phi \partial^{M}\phi \,-\, V(\phi)],
\label{S1}
\end{equation}
where $\phi(z)=m_{q}z+\sigma z^{3}$ or $\phi(z)=m_{q}z$ according to corresponding background solution, and $V(\phi)={\rm Tr}[V(X)]$.
Note that we assume $\sigma\!=\!0$ in AdSBH since this background corresponds to a high temperature deconfined phase
where the chiral symmetry is likely restored.

\subsection{Deformed thermal AdS}

Here we generalize the back-reacted AdS (tAdS) solution in \cite{Shock} to arbitrary $N_{f}$.
We assume the back-reacted AdS metric as
\begin{equation}
ds^{2}=e^{-2A(z)}(dt^{2}-d{\bf x}^{2})-dz^{2}/z^{2},
\label{metric}
\end{equation}
where $A(z)$ is an unknown function of the fifth coordinate $z$.
The equations of motion for (\ref{S1}) are given by
\begin{equation}
{1\over{2}}g_{MN}\left[-{R\over 2\kappa^{2}} - {N_{f}z^{2}\over{4}}(\phi')^{2}-V(\phi)  \right]\,+\, {1\over{2\kappa^{2}}}R_{MN}\,-\, {N_{f}\over{4}}\partial_{M}\phi\partial_{N}\phi\,=\,0\,,
\label{Eq1}
\end{equation}
\begin{equation}
{\partial V\over{\partial \phi}}\,=\, {N_{f}\over{2\sqrt{g}}}\partial_{M}(\sqrt{g}g^{MN}\partial_{N}\phi)\,,
\label{Eq2}
\end{equation}
where the prime denotes the derivative with respect to $z$.
$V[\phi(z)]$ will be denoted just as $V(z)$.
Then we can derive A(z) and V(z) from (\ref{Eq1}) for $\phi(z)=m_{q}z+\sigma z^{3}$ and these functions satisfy (\ref{Eq2}) automatically.
Two independent equations, one of which comes from the $00$ or $ii$ component of (\ref{Eq1}), and the other is from the $44$ component, are given by
\begin{align}
&{1\over{2\kappa^{2}}}[6z^{2}A''+6zA'-12z^{2}(A')^{2}]-V(z)-{N_{f}z^{2}\over{4}}(\phi')^{2}=0,
\label{Eq1-1}
\\
&{1\over{2\kappa^{2}}}[12z^{2}(A')^{2}]+V(z)={N_{f}z^{2}\over{4}}(\phi')^{2}.
\label{Eq1-2}
\end{align}
These equations are simplified into the following forms,
\begin{align}
&{1\over{2\kappa^{2}}}[6z^{2}A''+6zA']-{N_{f}z^{2}\over{2}}(\phi')^{2}=0,
\label{Eq1-3}
\\
&{1\over{2\kappa^{2}}}[3z^{2}A''+3zA'-12z^{2}(A')^{2}]-V(z)=0.
\label{Eq1-4}
\end{align}
We can check that these reduce to eq.(8) and eq.(9) in \cite{Shock} by setting $2\kappa^{2}=1$, $N_{f}=2$, flipping the sign of potential $V$, and changing the fifth coordinate $z \to y=\ln z $.
Then the back-reacted metric with $A(z)$ and the potential $V(z)$ are derived to be
\begin{align}
A(z)&=\ln z + 2\kappa^{2}N_{f}\left[{m_{q}^{2}\over{48}}z^{2} + {m_{q}\sigma\over{32}}z^{4} + {\sigma^{2}\over{48}}z^{6}\right],
\label{A(z)}
\\
V(z)&=-\Big[
{12\over{2\kappa^{2}}}+{3N_{f}\over{4}}(m_{q}z+\sigma z^{3})^{2}+2\kappa^{2}N_{f}^{2}\Big({m_{q}^{4}\over{48}}z^{4}+{m_{q}^{3}\sigma\over{8}}z^{6}
\nonumber\\
&\,\,\,\,\,\,\,\,\,\,\,\,\,\,\,\,\,\,\,\,\,+{5m_{q}^{2}\sigma^{2}\over{16}}z^{8}+{3m_{q}\sigma^{3}\over{8}}z^{10}+{3\sigma^{4}\over{16}}z^{12}\Big)
\Big],
\label{Vz1}
\end{align}
where the first term of $V(z)$ corresponds to the cosmological constant $-12/(2\kappa^2)$, and the second term corresponds to the negative mass term $-(3N_{f}/{4})(m_{q}z+\sigma z^{3})^{2}={\rm Tr}[-3X^{2}]$.
Thus the additional non-trivial potential $V_{\rm int}(z)$ is given by
\begin{equation}
V_{\rm int}(z)=
-2\kappa^{2}N_{f}^{2}\Big({m_{q}^{4}\over{48}}z^{4}+{m_{q}^{3}\sigma\over{8}}z^{6}+{5m_{q}^{2}\sigma^{2}\over{16}}z^{8}+{3m_{q}\sigma^{3}\over{8}}z^{10}+{3\sigma^{4}\over{16}}z^{12}\Big).
\label{Vint1}
\end{equation}
It is certain that these $A(z)$ and $V(z)$ reduce to those in Sec.3.1 of \cite{Shock} under the same conditions as stated above.

\subsection{Deformed AdS black hole}

Now we derive an AdSBH solution deformed by the back-reaction due to the scalar field.
We again start from the same $5$-dimensional action (\ref{S1}) with the following ansatz for a back-reacted AdSBH background
\begin{equation}
	ds^2=e^{-2A(z)}(f(z)dt^2-d{\bf x}^2)-dz^2/(z^2f(z))\;.
\end{equation}
With this setup, we can obtain a set of equations of motion from (\ref{Eq1}) as we do in the AdS case.
\begin{align}
	&\frac{1}{2\kappa^2}[f\{6z^{2}A''+6zA'-12z^{2}(A')^{2}\}+3z^2f'A']-V(z)-\frac{N_f z^2}{4}f(\phi')^2=0\,, \label{Eq2-1} \\
	&\frac{1}{2\kappa^2}[f\{6z^{2}A''+6zA'-12z^{2}(A')^{2}\}+f'(-z+7z^2A')-z^2f''] \nn \\
	&~~~~~~~~~~~~~~~~~~~~~~~~~~~~~~~~~~~~~~~~~~~~~~~~~~~~~~~~~~~-V(z)-\frac{N_f z^2}{4}f(\phi')^2=0\,, \label{Eq2-2} \\
	&\frac{1}{2\kappa^2}[12z^2f(A')^{2}-3z^2f'A']+V(z)-\frac{N_f z^2}{4}f(\phi')^{2}=0 \label{Eq2-3}
\end{align}
Each equation comes from the $00$, $ii$, and $44$ components of the equations of motion respectively.
These equations are re-combined into the simple forms,
\begin{align}
	&(-1+4\theta)f'-zf''=0\,, \label{Eq2-4} \\
	&6\theta'-\kappa^2 N_f z(\phi')^2=0\,, \label{Eq2-5} \\
	&2\kappa^2V(z)-3\{f(-4\theta^2+z\theta')+z\theta f'\}=0\,, \label{Eq2-6}
\end{align}
where $\theta(z)\equiv zA'(z)$.
As stated previously, we assume that $\phi(z)=m_q z$, \textit{i.e.} with vanishing chiral condensate for the black hole solution.
To solve the coupled equations, we first rewrite  (\ref{Eq2-5}) as
\begin{equation}
	\theta'(z)=\frac{\kappa^2 N_f}{6}m_{q}^{2}z\equiv 2C_{1}z~~~~\Rightarrow~~~\theta=C_{1}z^2+\tilde C_1 \;.
\end{equation}
Here  $\tilde C_1$ is fixed one by requiring that the our deformed solution should be the AdS  at the boundary, $z\to0$.
Inserting this $\theta(z)$ into (\ref{Eq2-4}), we obtain
\begin{eqnarray}
	&&f'(z)=-C_{2}z^3e^{2C_{1}z^2} \nonumber \\
	&&~~~~~\Rightarrow~~~~~f(z)=1-\frac{C_{2}}{8C_{1}^2}\{1-(1-2C_{1}z^2)e^{2C_{1}z^2}\}\;.
\end{eqnarray}
Also, from the behavior of $f(z)$ in the massless quark limit, we can choose $C_2=4/z_{h}^4$.
Here $z_{h}$ is the horizon of AdSBH without quark mass.
Now let's turn our attention to the potential $V[\phi(z)]$.
Above $\theta(z)$ and $f(z)$ should satisfy the remaining equation (\ref{Eq2-6}), so $V(z)$ is expressed as
\begin{eqnarray}
	V(z)&\!=\!&\frac{3}{2\kappa^2}\left[\left(-4+\frac{C_{2}}{2C_{1}^2}(1-e^{2C_{1}z^2})\right)\right. \nonumber \\
	 &&~~~~~~~\left.+C_{1}\left(-6+\frac{C_{2}}{4C_{1}^2}(3+e^{2C_{1}z^2})\right)z^2+C_{1}^2\left(-4+\frac{C_{2}}{2C_{1}^2}\right)z^4\right]\;.
\end{eqnarray}
This potential automatically satisfies the equation of motion (\ref{Eq2}) for the scalar field.
\begin{equation}
	\phi''+(\frac{f'}{f}-\frac{-1+4\theta}{z})\phi'-\frac{2}{N_fz^2f}\frac{\partial V}{\partial\phi}=0
\end{equation}
Let's summarize the newly found deformed AdSBH.
For convenience, we introduce the new parameter $z_{Q}^2\equiv1/(2C_1)=6/(\kappa^2 N_f m_{q}^2)$, then
\begin{align} \label{defBH}
 	&\phi(z)\equiv m_{q}z~, \nonumber \\
	&\theta(z)=1+\frac{1}{2}(\frac{z}{z_{Q}})^2~~~~~\Rightarrow~~~~A(z)=\ln z+\frac{1}{4}(\frac{z}{z_Q})^2~, \nonumber \\
	&f(z)=1-2(\frac{z_{Q}}{z_{h}})^4\left[1-\Big(1-(\frac{z}{z_{Q}})^2\Big)e^{(z/z_{Q})^2}\right]
\end{align}
and
\begin{eqnarray} \label{VzBH}
	V[\phi(z)]&\!=\!&\frac{3}{2\kappa^2}\left[-4\left(1-2(\frac{z_{Q}}{z_{h}})^4\right)(1+\frac{\phi^4}{4z_{Q}^4 m_{q}^4})-8(\frac{z_{Q}}{z_{h}})^4e^{\phi^2/(z_{Q}^2 m_{q}^2)}\right. \nonumber \\
	&&~~~~~~~~~~~~~~~~~~~~~~~~~\left.-\left(3-2(\frac{z_{Q}}{z_{h}})^4(3+e^{\phi^2/(z_{Q}^2 m_{q}^2)})\right)\frac{\phi^2}{z_{Q}^2 m_{q}^2}\right]\;.
\end{eqnarray}
This solution is asymptotically AdS and reduces to the original AdSBH in the limit $m_q\to 0$.
The Hawking temperature of this black hole is
\begin{equation} \label{tempAdSBH}
	T=\frac{1}{\pi z_H}(\frac{z_H}{z_h})^4 e^{\frac{3}{4}(\frac{z_H}{z_Q})^2}
\end{equation}
where $z_H$ is the horizon of this solution satifying $f(z_H)=0$.
The concrete form of $z_H$ is given by 
\begin{equation}
z_H = z_Q\sqrt{1+\textrm{ProductLog}\,\Big[\frac{(z_h/z_Q)^4-2}{2e}\Big]}
\end{equation}
where $\textrm{ProductLog}\,[a]$ stands for a solution for $x$ in the eqeuation $xe^x\!=\!a$.
We note here that for small $m_q$, $z_H$ is approximately equal to $z_h$.

\section{Hawking-Page transition}
In this section, we do a Hawking-Page type analysis using the back-reacted backgrounds of tAdS and AdSBH discussed in the previous section.
Primary goals here are to remove the deconfined phase at low temperature, unwanted by-product of  \cite{Kim:2007em},
and to see the quark mass dependence of the critical
temperature for the deconfinement transition.
We work in the hard wall model~\cite{EKSS, PR}, where the AdS space is compactified such that $\epsilon<z<z_{0}$ with $\epsilon\to0$.
$z_{0}=z_{\rm IR}$ is an IR cut-off which models the confinement in QCD and is fixed by the phenomenological quantities such as the $\rho$-meson mass.
We will derive the on-shell action densities, using the Euclidean bulk action,
\begin{equation} \label{S3}
S_{E}=S_{grav}+S_{mat}=\int d^{5}x \sqrt{g}~[-{1\over{2\kappa^{2}}}R\,+\,{\rm Tr}(\partial_{M} X\partial^{M}X\,+\,V(X))]\,,
\end{equation}
 for the back-reacted tAdS and for the back-reacted AdSBH in Euclidean space and compare them to calculate
 the critical temperature for various $m_{q}$ and $N_{f}$.
In the discussion of the phase transition,  it is enough to consider the above Euclidean action (\ref{S3}) 
since the gauge fields do not contribute to the action density.

In calculation of the action density, we use the time periodicity in the integration along the Euclidean time coordinate $\tau$.
The time periodicity of AdSBH is determined as $\beta=1/T$, while that of tAdS is fixed by comparing two geometries at the UV cut-off and given by
\begin{equation} \label{AdStimeP}
\beta'=\frac{1}{T}\sqrt{f(\epsilon)}\,e^{2\kappa^2 N_{f}(\frac{m_{q}\sigma}{32}\epsilon^4+\frac{\sigma^2}{48}\epsilon^6)}
\end{equation}
where $T$ stands for the Hawking temperature of AdSBH in (\ref{tempAdSBH}) and $f(z)$ is given by the third equation in (\ref{defBH}).

\subsection{Thermal AdS}
Let us begin with the on-shell action of tAdS.
The deformed metric is ~\cite{Shock}
\begin{align}
ds^{2}\,&=\, e^{-2A(z)}(d\tau^{2}+d{\bf x}^{2})\,+\,dz^{2}/z^{2},
\nonumber\\
A(z)&=\ln z+2\kappa^{2}N_{f}\left[{m_{q}^{2}\over{48}}z^{2} + {m_{q}\sigma\over{32}}z^{4} + {\sigma^{2}\over{48}}z^{6}\right]\,.
\label{mettAdS}
\end{align}
Then the corresponding Ricci scalar is given by
\begin{align}
R\,&=\, -4(-2zA'+5z^{2}(A')^{2}-2z^{2}A'')
\nonumber\\
&=\, -20-2\kappa^{2}N_{f}\Big( m_{q}^{2}z^{2}+m_{q}\sigma z^{4}-\sigma^{2}z^{6} \Big)
-(2\kappa^{2})^{2}N_{f}^{2}\Big( {5\over{144}}m_{q}^{4}z^{4}
\nonumber\\
&\,\,\,\,\,\,\,\,\,+{5\over{24}}m_{q}^{3}\sigma z^{6}+{25\over{48}}m_{q}^{2}\sigma^{2}z^{8}+{5\over{8}}m_{q}\sigma^{3}z^{10}+{5\over{16}}\sigma^{4}z^{12} \Big).
\end{align}
We can easily confirm that this Ricci scalar results in that of the usual AdS, $R=-20$, by setting both the quark mass and the chiral condensate zero.
The gravitational on-shell action for the back-reacted tAdS is given by
\begin{align}
S_{grav}\,&=\, -{1\over{2\kappa^{2}}}\int d^{5}x\,\sqrt{g} \left( R\,+\, 12 \right)\,.
\end{align}
Now we calculate the action density $V \equiv S_{E}/ \int d^3x$ for the gravitational sector:
\begin{equation}
V_{1g} = {4\over{\kappa^{2}}}\beta'
\int^{z_{0}}_{\epsilon}{dz\over{z^{5}}}e^{-B(z)}\Big( -{R\over{8}}\,-\,{3\over{2}} \Big)
\label{V1g+m}
\end{equation}
where
\begin{equation}
B(z)=2\kappa^{2}N_{f}\left[{m_{q}^{2}\over{12}}z^{2} + {m_{q}\sigma \over{8}}z^{4} + {\sigma^{2}\over{12}}z^{6}\right]
\end{equation}
and the time peridocity $\beta'$ written in (\ref{AdStimeP}) are used.

The matter part of the action in Euclidean space on the back-reacted tAdS background is given by
\begin{align}
S_{mat}&= \int d^5x \sqrt{g}~
	{\rm Tr} \left[ \, g^{zz}(\partial_z X_0)^2 \,- \, 3 X_0^2+V_{\rm int}(X_{0}) \right]
\nonumber\\
&=   \int d^5x {e^{-B(z)}\over{z^{5}}}
	\left[ \, z^{2}(\partial_z X_0)^2 \,- \, 3 X_0^2+V_{\rm int}(X_{0}) \right].
\label{Smatter1}
\end{align}
For simplicity we drop the trace ${\rm Tr}$ in the second line.
(As we will see, this trace just gives a factor $N_{f}$.)
To derive the corresponding on-shell action,
we perform a partial integral in (\ref{Smatter1}) as
\begin{align}
S_{mat}&=  \int d^{5}x
	\left[ -X_{0}  \partial_{z}  \Big(  {e^{-B}\over{z^{3}}} \partial_{z} X_{0} \Big) -  {3e^{-B}\over{z^{5}}} X_{0}^{2} + {e^{-B}\over{z^{5}}}V_{\rm int}(X_{0}) \right]+\int d^{4}x \left[ {e^{-B}\over{z^{3}}}X_{0}\partial_{z} X_{0} \right]^{z_{0}}_{\epsilon}
\nonumber\\
&=
-\int d^{5}x  X_{0}\Big[ {\rm E.O.M.} + {e^{-B}\over{2z^{5}}}{\partial V_{\rm int}(X_{0})\over{\partial X_{0}}} - {e^{-B}\over{z^{5}}}{V_{\rm int}(X_{0})\over{X_{0}}} \Big]
\nonumber\\
&~~~~~~~~~~~~~~~~~~~~~~~~~~~~~~~~~~~~~~~~~~~~~~~~~~~~~~~~~~~~~~~~~+\, \int d^{4}x \left[ {e^{-B}\over{z^{3}}}X_{0}\partial_{z} X_0 \right]^{z_{0}}_{\epsilon}.
\end{align}
Here E.O.M. stands for the equation of motion of $X_{0}(z)$, which is given by
\begin{equation}
{\rm E.O.M.}:\,\, \partial_{z}\Big( {e^{-B}\over{z^{3}}} \partial_{z} X_{0}   \Big) + {3e^{-B}\over{z^{5}}} X_{0} -{e^{-B}\over{2z^{5}}}{\partial V_{\rm int}(X_{0})\over{\partial X_{0}}}=0 .
\end{equation}
Then we obtain  the on-shell action for the scalar field on the back-reacted tAdS as
\begin{equation}
S_{mat}=\int d^{4}x \left[ {e^{-B}\over{z^{3}}}X_{0}\partial_{z} X_0 \right]^{z_{0}}_{\epsilon}-\int d^{5}x  X_{0}\Big[ {e^{-B}\over{2z^{5}}}{\partial V_{\rm int}(X_{0})\over{\partial X_{0}}} -{e^{-B}\over{z^{5}}}{V_{\rm int}(X_{0})\over{X_{0}}} \Big].	
\end{equation}
From this on-shell action, we derive the action density for the matter field on the tAdS background as follows,
\begin{equation} \label{acD1m}
V_{1m}(\epsilon)=\beta'\left[{e^{-B}\over{z^{3}}}X_{0}\partial_{z} X_0 \right]^{z_{0}}_{\epsilon}-\beta'\int^{z_{0}}_{\epsilon}dz {e^{-B}\over{2z^{5}}}\Big[X_{0}{\partial V_{\rm int}(X_{0})\over{\partial X_{0}}}-2V_{\rm int}(X_{0})\Big]\,.
\end{equation}
It is clear that the potential $V_{\rm int}$ yields an additional non-boundary term.
As discussed in Sec.2, the real scalar solution $X_{0}$ keeps the same form for all the region $\epsilon<z<z_{0}$ as
\begin{equation}
X_{0}(z)\,=\,{{\bf 1}_{N_{f}\times N_{f}}\over{2}}(m_{q}z+\sigma z^{3}).
\end{equation}
Thus, the first term in (\ref{acD1m}) denoted as $V_{1m}^{\rm (1)}$ is easily calculated as follows,
\begin{align}
V_{1m}^{\rm (1)}(\epsilon)={N_{f}\over{4}}\beta'\Big[e^{-B(z_{0})}\Big( {m_{q}^{2}\over{z_{0}^{2}}}+3\sigma^{2}z^{2}_{0}+4m_{q}\sigma \Big)-e^{-B(\epsilon)}\Big({m_{q}^{2}\over{\epsilon^{2}}}+3\sigma^{2}\epsilon^{2}+4m_{q}\sigma \Big)\Big].
\label{V1m1}
\end{align}
Here we carry out the trace which gives a factor $N_{f}$.
Next we derive $V_{\rm int}(z)$ for the back-reacted tAdS in order to calculate the second term in (\ref{acD1m}) denoted as $V_{1m}^{(2)}$.
The total potential $V(z)$ is given by (\ref{Vz1}),
and the additional potential $V_{\rm int}(z)$ is given by (\ref{Vint1}).
By substituting this $V_{\rm int}(z)$ and $X_{0}(z)$ or $\phi(z)$, the second term $V_{1m}^{(2)}$ in (\ref{acD1m}) is calculated as,
\begin{align}
V_{1m}^{(2)}(\epsilon) &=-\beta'\int^{z_{0}}_{\epsilon} dz  {e^{-B}\over{2z^{5}}}\Big[{X_{0}\over{\partial_{z}X_{0}}} \partial_{z} V_{\rm int}(z) -2V_{\rm int}(z) \Big]
\nonumber\\
&=\beta'\int^{z_{0}}_{\epsilon} dz  {e^{-B}\over{2z^{5}}}{\kappa^{2}N_{f}^{2}\over{2}}\Big[ {m_{q}^{4}\over{6}}z^{4}+{4m_{q}^{3}\sigma\over{3}}z^{6}+{7m_{q}^{2}\sigma^{2}\over{2}}z^{8}
\nonumber\\
&~~~~~~~~~~~~~~~~~~~~~~~~~~~~~~~~~~~~~~~+4m_{q}\sigma^{3}z^{10}+{3\sigma^{4}\over{2}}z^{12} \Big].
\end{align}

\subsection{AdS black hole}

Now we consider the deformed AdSBH.
The metric is again
\begin{align}
ds^{2}\,&=\, e^{-2A(z)}(f(z)d\tau^{2}\,+\,d{\bf x}^{2})\,+\,dz^{2}/(z^{2}f(z)),
\nonumber\\
A(z)\,&=\, \ln z+\frac{1}{4}(\frac{z}{z_Q})^2,
\nonumber\\
f(z)\,&=\, 1-2(\frac{z_{Q}}{z_{h}})^4\left[1-\Big(1-(\frac{z}{z_{Q}})^2\Big)e^{(z/z_{Q})^2}\right],
\label{metAdSBH}
\end{align}
with $z_{Q}^{2}=6/(\kappa^{2}N_{f}m_{q}^{2})$.
The Ricci scalar of the background is given by,
\begin{align}
R\,&=\, -4(-2zfA' +5z^{2}f(A')^{2}-2z^{2}fA'' )-zf'+9z^{2}f'A' -z^{2}f''
\nonumber\\
&=\, -20-2\kappa^{2}N_{f}m_{q}^{2}z^{2}-(2\kappa^{2})^{2}{5N_{f}^{2}\over{144}}m_{q}^{4}z^{4}+\,\cdots\, ,
\end{align}
where the second line shows the common terms with the tAdS case, which cancel the divergence of the action densities.
Then, the gravitational action density for the back-reacted AdSBH is given by
\begin{equation}
V_{2g} = {4\over{\kappa^{2}}}\beta
\int^{min(z_{0},z_{H})}_{\epsilon}{dz\over{z^{5}}}e^{-B(z)}\Big( -{R\over{8}}\,-\,{3\over{2}} \Big)
\label{V2g+m}\, ,
\end{equation}
where $B(z)=2\kappa^{2}N_{f}m_{q}^{2}z^{2}/12$ and $\beta$ is used as the time periodicity of AdSBH solution.

The matter part of the Euclidean action on the back-reacted AdSBH background is given by
\begin{align}
S_{mat}&= \int d^5x \sqrt{g}~
	{\rm Tr} \left[ \, g^{zz}(\partial_z X_0)^2 \,- \, 3 X_0^2+V_{\rm int}(X_{0}) \right]
\nonumber\\
&=   \int d^5x {e^{-B(z)}\over{z^{5}}}
	\left[ \, f z^{2}(\partial_z X_0)^2 \,- \, 3 X_0^2+V_{\rm int}(X_{0}) \right].
\label{Smatter2}
\end{align}
Integration by part yields the on-shell action for matter sector.
\begin{align}
S_{mat}&=  \int d^{5}x
	\left[ -X_{0}  \partial_{z} \Big({f e^{-B}\over{z^{3}}} \partial_{z} X_{0} \Big) -  {3e^{-B}\over{z^{5}}} X_{0}^{2} + {e^{-B}\over{z^{5}}}V_{\rm int}(X_{0}) \right]
\nonumber\\
&\,\,\,\,\,\,\,\,\,\,\,\,\,\,\,\,\,\,\,\,\,\,\,\,\,\,\,\,\,\,\,\,\,\,\,\,\,\,\,\,\,\,\,\,\,\,\,\,\,\,\,\,\,\,\,\,\,\,\,\,\,\,\,\,\,\,+ \int d^{4}x \left[ {f e^{-B}\over{z^{3}}}X_{0}\partial_{z} X_{0} \right]^{{\rm min}(z_{0},z_{H})}_{\epsilon}
\nonumber\\
&=
-\int d^{5}x  X_{0}\Big[ {\rm E.O.M.} +  {e^{-B}\over{2z^{5}}}{\partial V_{\rm int}(X_{0})\over{\partial X_{0}}} -{e^{-B}\over{z^{5}}}{V_{\rm int}(X_{0})\over{X_{0}}} \Big]	
\nonumber\\
&\,\,\,\,\,\,\,\,\,\,\,\,\,\,\,\,\,\,\,\,\,\,\,\,\,\,\,\,\,\,\,\,\,\,\,\,\,\,\,\,\,\,\,\,\,\,\,\,\,\,\,\,\,\,\,\,\,\,\,\,\,\,\,\,\,\,+ \int d^{4}x \left[ {f e^{-B}\over{z^{3}}}X_{0}\partial_{z} X_0 \right]^{{\rm min}(z_{0},z_{H})}_{\epsilon}
\end{align}
where
\begin{equation}
{\rm E.O.M.}:\,\, \partial_{z}\Big( {f e^{-B}\over{z^{3}}} \partial_{z} X_{0}   \Big) + {3e^{-B}\over{z^{5}}} X_{0} -{e^{-B}\over{2z^{5}}}{\partial V_{\rm int}(X_{0})\over{\partial X_{0}}}=0.
\end{equation}
From this on-shell action, after integrating out the volume of $\mathbb{R}^3$, the matter sector of action density for the back-reacted AdSBH is
\begin{align}
V_{2m}(\epsilon) &= \beta \left[ {f e^{-B}\over{z^{3}}}X_{0}\partial_{z} X_0 \right]^{{\rm min}(z_{0},z_{H})}_{\epsilon}
\nonumber\\
&~~~~~~~~~~~~~- \int ^{{\rm min}(z_{0},z_{H})}_{\epsilon}dz  {e^{-B}\over{2z^{5}}}\Big[X_{0} {\partial V_{\rm int}(X_{0})\over{\partial X_{0}}} -2V_{\rm int}(X_{0}) \Big].
\label{acD2m}
\end{align}
Here we take $X_{0}(z)=({\bf 1}_{N_{f}\times N_{f}}/2)m_{q}z$.
Then, we can calculate the boundary term in (\ref{acD2m}).
\begin{equation} \label{V2m1}
V_{2m}^{\rm (1)}(\epsilon )= {N_{f}\over{4}}\beta\Big[f({\rm min}(z_{0},z_{H})) e^{-B({\rm min}(z_{0},z_{H}))} {m_{q}^{2}\over{{\rm min}(z_{0},z_{H})^{2}}}- \left( {m_{q}^{2}\over{\epsilon^{2}}}-{\kappa^{2}N_{f}m_{q}^{4}\over{6}} \right) \Big]
\end{equation}
We have taken the limit $\epsilon\to 0$ and neglected all terms with $\epsilon$, except the divergent one $m^2_q/\epsilon^2$.
To calculate the remaining part, we need to know the form of $V_{\rm int}(z)$ for the back-reacted AdSBH.
The total potential $V(z)$ for the AdSBH is given by replacing $\phi(z)=m_q z$ in (\ref{VzBH}), which contains the cosmological constant term $-12/(2\kappa^{2})$ and the negative-mass term $-3N_{f}m_{q}^{2}z^{2}/4$.
Then the additional potential $V_{\rm int}(z)$ is obtained by subtracting these terms from $V(z)$.
\begin{equation}
V_{\rm int}(z)=
-2\kappa^{2}{N_{f}^{2}m_{q}^{4}\over{48}}z^{4}-\frac{N_f m_q^2}{12z_h^4}z^6+\,\cdots\,.
\end{equation}
This $V_{\rm int}(z)$ contains the common term with the tAdS case, which leads to cancellation of the divergence of the action density.
By substituting $V_{\rm int}(z)$ and $X_{0}(z)$, we can write down the second term $V_{2m}^{(2)}$ in (\ref{acD2m}).
\begin{equation}
V_{2m}^{(2)}(\epsilon) =-\beta\int^{{\rm min}(z_{0},z_{H})}_{\epsilon}dz{e^{-B}\over{2z^{5}}}
\Big[{X_{0}\over{\partial_{z}X_{0}}} \partial_{z} V_{\rm int}(z) -2V_{\rm int}(z) \Big]
\end{equation}

\begin{figure}
\centering
\includegraphics[width=0.5 \textwidth]{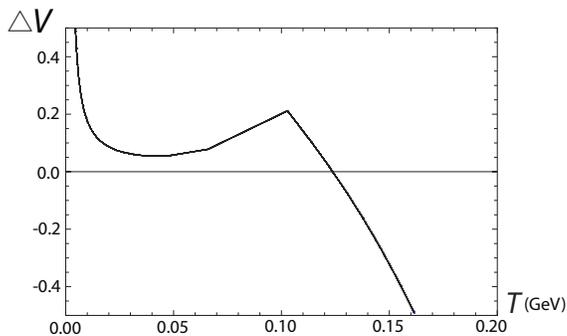}
\caption{The difference of the action densities $\triangle V(m_{q}=2.4~{\rm MeV}, N_{f}=2, N_{c}=3)$  as a function of temperature.}
\label{DAD}
\end{figure}

\subsection{Critical temperature}

Finally, we can write down the total difference of action density as follows
\begin{equation}
\triangle V \,\,=\,\,(V_{2g}+V_{2m})-(V_{1g}+V_{1m}).
\end{equation}
We evaluate this quantity numerically and confirm that the result is independent of $\epsilon$.
$\triangle V$ as a function of temperature is depicted in Fig.\ref{DAD}.
It is obvious that there is no unphysical phase at low temperature which appears in \cite{Kim:2007em}.
By setting $N_f=0$, our $\triangle V$ reduces to that in \cite{Herzog}.

\begin{figure}
\centering
\includegraphics[width=0.5 \textwidth]{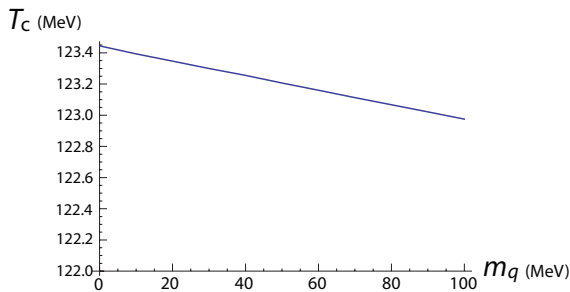}
\caption{The quark mass dependence of the critical temperature for $N_{f}=2, N_{c}=3$. }
\label{Tc-mq}
\end{figure}

\begin{figure}
\centering
\includegraphics[width=0.5 \textwidth]{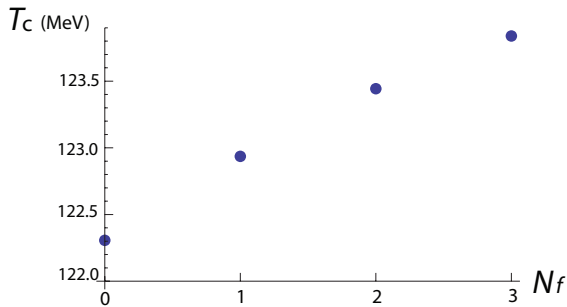}
\caption{The $N_{f}$ dependence of the critical temperature for $m_{q}=0, N_{c}=3$. }
\label{Tc-Nf}
\end{figure}

Next, we study the quark mass dependence of the critical temperature.
The result for $N_{f}=2, N_{c}=3$ is depicted in Fig.\ref{Tc-mq}.
$T_{c}$ decreases as the quark mass $m_{q}$ getting large.
We can also see the $N_{f}$ dependence of the critical temperature.
The result for $m_{q}=0, N_{c}=3$ is depicted in Fig.\ref{Tc-Nf}.
$T_{c}$ increases with $N_{f}$ getting large.

\section{Summary}

In this paper we have studied the confinement/deconfinement transition in holographic QCD considering the back-reaction of a bulk scalar field.
We first obtained a deformed AdS black hole solution due to a finite quark mass.
Then, through the Hawking-Page analysis with the back-reacted geometry,
we study the flavor number $N_f$ and finite quark mass dependence of the critical temperature
of the QCD confinement/deconfinement transition.

We observed that $T_{\rm c}$ decreases with the increasing quark mass $m_{q}$, 
while it increases with $N_{f}$.
These results are not consistent with the lattice QCD both in $m_{q}$ and $N_{f}$ dependence of $T_{\rm c}$, for instance see~\cite{KLP}.
We remark here that in the previous study without back-reaction~\cite{Kim:2007em} the results on   $N_{f}$ dependence
was consistent with the lattice QCD result.
This may imply the importance of including complete $1/N_{c}$ corrections in holographic QCD.
In ~\cite{Kim:2007em}, only matter part of the bulk action was considered, and so only a single source of $1/N_{c}$ corrections
is included.
In our work we considered the matter action and back-reaction.
There are, however, some other sources of $1/N_{c}$ corrections such as terms with higher order derivatives.
Therefore we conclude that to compare results from a holographic QCD study with those from lattice QCD or experiments,
we have to include complete $1/N_{c}$ corrections, not part of them.

\acknowledgments
T.M. thanks Koji Hashimoto and Elias Kiritsis for useful comments.
Y.K. and T.M. are grateful to Kenji Fukushima for helpful discussions.
Y.K. and I.J.S. acknowledge the Max Planck Society(MPG), the Korea Ministry of Education,
Science, Technology(MEST), Gyeongsangbuk-Do and Pohang City for the support
of the Independent Junior Research Group at the Asia Pacific Center for
Theoretical Physics(APCTP).
T.M. is supported by Grand-in-Aid for the Japan Society for Promotion of Science (JSPS) Research Fellows (No.21-1226).

\end{document}